\documentclass[%
 aip,
 amsmath,amssymb,
 reprint,%
]{revtex4-1}

\usepackage{graphicx}
\usepackage{dcolumn}
\usepackage{bm}

\usepackage[utf8]{inputenc}
\usepackage[T1]{fontenc}
\usepackage{mathptmx}
\usepackage{physics}

\usepackage{xcolor}
\usepackage{soul}

\definecolor{orange}{HTML}{FF7F00}
\definecolor{cyan}{HTML}{0066CC}
\definecolor{bordeaux}{HTML}{8B008B}
\definecolor{red}{HTML}{FF0000}

\newcommand{\bb}[1]{\boldsymbol{#1}}

\newcommand{\ba}[1]{\bar{#1}}
 
\newcommand{\Hm}{\bb{H}}
\newcommand{\Sm}{\bb{S}}
\newcommand{\Pm}{\bb{P}}
\newcommand{\Um}{\bb{U}}

\newcommand{\Gm}{\bb{G}}
\newcommand{\Am}{\bb{A}}

\newcommand{\Hp}{\bb{H}^{\prime}}

\newcommand{\Ht}{\Hm}
\newcommand{\St}{\Sm}

\newcommand{\Hb}{\ba{\bb{H}}}
\newcommand{\Sb}{\ba{\bb{S}}}

\newcommand{\Sigb}{\ba{\bb{\Sigma}}}
\newcommand{\gb}{\ba{\bb{g}}}
\newcommand{\Gb}{\ba{\bb{G}}}

\newcommand{\Rv}{\bb{R}}
\newcommand{\kv}{\bb{k}}

\newcommand{\Ov}{\bb{0}}

\usepackage{amsmath}
\usepackage{multirow}
\usepackage{tikz}
\usetikzlibrary{matrix}

\begin{document}

\preprint{AIP/123-QED}

\title{Smart local orbitals for efficient calculations within density functional theory and beyond}


\author{G.~Gandus}
 \affiliation{Empa, Swiss Federal Laboratories for Materials Science and Technology, {\"U}berlandstrasse 129, 8600, D{\"u}bendorf, Switzerland}
 \email{ggandus@ethz.ch}
\author{A.~Valli}
 \affiliation{Institute for Theoretical Physics, TU Wien, Wiedner Hauptstrasse 8-10, 1040 Vienna, Austria}
\author{D.~Passerone}
 \affiliation{Empa, Swiss Federal Laboratories for Materials Science and Technology, {\"U}berlandstrasse 129, 8600, D{\"u}bendorf, Switzerland}
\author{R.~Stadler}
 \affiliation{Institute for Theoretical Physics, TU Wien, Wiedner Hauptstrasse 8-10, 1040 Vienna, Austria} 

\date{\today}

\begin{abstract}
Localized basis sets in the projector augmented wave formalism allow for 
computationally efficient calculations within density functional theory (DFT). 
However, achieving high numerical accuracy requires an extensive basis set, 
which also poses a fundamental problem for the interpretation of the results. 
We present a way to obtain a reduced basis set of atomic orbitals
through the subdiagonalization of each atomic block of the Hamiltonian. 
The resulting local orbitals (LOs) inherit the information of the local crystal field. 
In the LO basis, it becomes apparent that the Hamiltonian is nearly block-diagonal, 
and we demonstrate that it is possible to keep only a subset of relevant LOs 
which provide an accurate description of the physics around the Fermi level. 
This reduces to some extent the redundancy of the original basis set, 
and at the same time it allows one to perform post-processing of DFT calculations, 
ranging from the interpretation of electron transport 
to extracting effective tight-binding Hamiltonians, 
very efficiently and without sacrificing the accuracy of the results. 
\end{abstract}

\maketitle


\section{Introduction}\label{sec:intro}

The recent developments in the fabrication and the characterization of low-dimensional materials  
attracted a lot of interest both from the point of view of fundamental research, 
providing a relatively simple platform for exploring exotic quantum effects, 
and for the potential they hold for applications. 
Theory plays an important role in the interpretation of the experimental data, 
for its ability to rationalize complex phenomena 
in terms of fundamental physical processes. 
At the same time, numerical predictions can also guide the experiments 
towards optimal choices of materials and properties 
in a synergistic effort to improve the performance of devices. 
For this reason, it is important for theoretical simulations to be as accurate as possible. 
In this regard, density functional theory (DFT) has established itself as the standard approach 
to investigate the electronic properties of materials at the single-particle level. 
Moreover, the theory of electron transport in the framework of DFT,~\cite{cuevasME} 
within a non-equilibrium Green's function (NEGF) formalism,~\cite{stefanucciNEMBTQS} 
has been the reference approach to shed light on the behaviour of nanoscale devices. 
However, a systematic control of the numerical accuracy 
to achieve quantitative numerical predictions is still challenging.~\cite{eversRMP92}

In addition to the approximate nature of the DFT exchange-correlation functional, a properly chosen basis 
is necessary to obtain results with a reasonable accuracy. 
In practice, one needs to consider an extensive basis set to ensure 
a sufficient flexibility (namely, a large enough number of variational coefficients) 
for a correct description of the electronic wave function.
This poses a fundamental problem for the interpretation of the result of the calculation.  
To gain an enhanced physical understanding, 
it can be useful to 
extract a reduced set of intelligible atomic-like orbitals.
For instance, it comes natural to interpret chemical bonding in terms of localized orbitals 
and the mechanisms underlying electron transport 
in terms of atomic or molecular orbitals.~\cite{xinJCP146} 
Moreover, methods that rely on the evaluation of the Green's function, such as NEGF for transport,  
or including many-body corrections, e.g., within GW or dynamical mean-field theory~\cite{georgesRMP68} (DMFT) 
in combination with DFT,~\cite{kotliarRMP78,heldAP56,schuelerEPJBST226,tomczakEJPBST226} 
become impractical with extensive basis sets. 

In this paper, we take a first step towards a deterministic approach to extract a minimal basis 
from a linear combination of atomic orbitals (LCAO) calculation. 
From the corresponding tight-binding Hamiltonian and overlap matrices, 
we demonstrate that we can perform calculations of band structure, 
charge occupation analysis, and quantum transport 
with a significant reduction of the numerical costs without sacrificing the accuracy of the results. 

Extensive work has been done in this direction in the framework of plane-wave basis sets. 
The Wannier function (WF) approach identifies a set of spatially-localized orbitals 
via a unitary transformation of the Kohn-Sham wave functions, 
whose parameters can be obtained through an iterative minimization of a certain functional. 
In the case of Boys' localization~\cite{boys_construction_1960} 
the functional is the sum of the quadratic spread of the localized molecular orbitals. 
Other examples include maximizing the Coulomb self-interaction of the orbitals
(Edmiston-Ruedenberg)~\cite{edmiston_localized_1963}, 
the density overlap of the orbitals (Von Niessen)~\cite{von_niessen_density_1972}, 
or the sum of the squares of the Mulliken atomic charges (Pipek-Mezey)~\cite{pipek_fast_1989}. 
An important contribution in this regard was the introduction of maximally localized Wannier functions (MLWF) 
by Marzari and Vanderbilt,~\cite{marzari_maximally_1997,marzariRMP84} 
which paved a unique way to postprocess electronic structure calculations. 
Notwithstanding the usefulness of MLWFs, the construction of WFs 
is far from trivial.~\cite{strangeJCP128,thygesen2005molecular} 
The problem arises from determining a sensible choice of initial trial wavefunctions 
and a target band manifold that are both required for the iterative scheme 
and which uniquely define the resulting WFs. 
Furthermore, the center and the form of the WFs is not known a-priori.
For transport, where the system is partitioned into semi-infinite leads and a channel, this implies that particular care must be taken when extracting the couplings between the leads and the central region from separate electronic calculations, as the bases in the different regions are not guaranteed to be equivalent.

Other related methods, 
like the quasiatomic orbital (QO) scheme,~\cite{lu_molecule_2004,qian_quasiatomic_2008,qian_quasiatomic_2010} 
extract a set of WFs that are constructed with the criterion that they are maximally similar 
to a pre-selected set of atomic orbitals with defined symmetries. 
While a closed-form solution for the QOs exists, such orbitals are guaranteed to be centered 
at the atomic positions, but they are not maximally localized. 
Therefore, care must be taken when constructing a corresponding 
tight-binding Hamiltonian.~\cite{qian_quasiatomic_2008} 
In this context, we also mention the mode space approximation,~\cite{mil2012equivalent,ducry2020hybrid} 
which is a variational method to construct a reduced basis set of transverse modes 
reproducing the physical states of a periodic system within an arbitrary energy window. 
In the framework of quantum transport, this allows to reproduce the bandstructure of periodic leads.

Here we take a different approach. We demonstrate that it is possible to generate 
a reduced set of localized orbitals that inherit the information of each atom's environment, 
directly from the LCAO Hamiltonian, instead of performing a projection of the Kohn-Sham states. 
We refer to this basis set as local orbitals (LOs).
The LOs are constructed for any atom in the system 
through a subdiagonalization of the Hamiltonian block of its AOs.
This procedure yields a set of LOs 
which are atomic-like functions and are by construction 
i) atom-centered, and ii) orthogonal within the same atom (but not among different atoms). 
Furthermore, the LO representation can coexist with the original AO one, 
in the sense that one can subdiagonalize only a subset of atoms within the Hamiltonian of the system. 
This is useful if one is particularly interested in a limited part of a system,  
such as a molecular bridge in a quantum junction, or an adsorbate on a substrate. 

We mention that our approach does not aim at dividing the electronic density of the system into atomic charges as in Bader's partitioning scheme.~\cite{bader_1990} Here, we operate at the wave function level, 
obtaining atomic-like orbitals which have a clear physical interpretation and also 
allow one to discard less relevant degrees of freedom. 
However, in the context of charge population analysis as the one developed by Mulliken's,~\cite{mulliken_1955} the LOs can be exploited to obtain an orbital-resolved partial charge analysis since, contrary to the AOs, they take into account the atomic crystal field environment.

We shall demonstrate that in the LO basis the Hamiltonian can be partitioned 
into sub-blocks which are, to a first approximation, independent. 
All LOs with a given character but centered on different atoms 
are grouped in the same sub-block. 
Indeed a well-defined procedure can be adopted to identify a reduced set of LOs 
that are kept, while the rest can be discarded in order to define effective tight-binding Hamiltonians.  
The proposed separation of degrees of freedom thus allows to, e.g., compute transport properties 
around the Fermi level with a substantially reduced computational effort, 
or disentangle overlapping bands. 
Electronic and transport properties computed within this approximation are shown to be 
virtually identical to the ones obtained with the original basis set. 
We refer to this method as cut-coupling, as discussed in Sec.~\ref{sec:cut}. 
One can also take a further step, which reintroduces the influence of the discarded orbitals 
as embedding by means of a self-energy matrix, as discussed in Sec.~\ref{sec:emb}.


We demonstrate our method by performing benchmark calculations against the full LCAO basis set. 
The benchmark is achieved by comparing the transmission function 
of two nanoscale contacts and the band structure of three 
periodic systems consisting of a single monolayers. 
The reference systems for transport calculations are 
i) a planar organic polyacene (PA) junction and 
ii) a benzene-diamine (BDA) molecule bridging Au electrodes, 
which are representative of a wide class of currently explored all-carbon nanodevices (the former), 
and typical single-molecule junctions 
contacted by metallic leads (the latter). 
For the band electronic structure, we have chosen 
graphene, hexagonal boron nitride, and molybdenum disulfide, 
which covers classes of materials with very different chemical and physical properties. 

In all cases, the results with the LO basis set 
are shown to be virtually identical to the ones obtained with the full basis set. 
At this point, we remark that our accuracy tests are preformed against the LCAO result, 
which is our reference. 
We do not explicitly compare our method against WFs, but it was shown~\cite{strangeJCP128} 
that transport calculations within the LCAO framework 
using the double-$\zeta$ polarized basis agree also very well with the results 
obtained with MLWF calculations. 

The LOs are also appealing in the context of many-body calculations beyond DFT, 
where one separates an active space, i.e., a minimal set of LOs with the same character, 
that is intended to describe the energy range of interest, 
from an embedding 
that reintroduces the
influence of the other orbitals via a self-energy matrix approach.
However, this is beyond the scope of the present work, and will be explored elsewhere. 
Finally, we also note that our implementation assumes 
that the electron spin degeneracy is not lifted, but it could be generalized to treat spin-orbitals. 
This would allow to address also quantum junction displaying 
spin-selective transport properties.~\cite{valliNL18,valliPRB100,guo_2019,Zollner_2020} 

The paper is organized as follows. In Sec.~\ref{sec:frame}, we discuss the theoretical and mathematical framework
to obtain the LOs. 
In Sec.~\ref{sec:gpaw_details} we give the details of the numerical calculations performed in this work. 
In Secs.~\ref{sec:appl_transport}, ~\ref{sec:appl_electr}, and ~\ref{sec:bands} 
we present the results of the benchmark calculations for the reference systems. 
Finally, Sec.~\ref{sec:outlook} contains a summary and an outlook.

\section{Theoretical Framework}\label{sec:frame}
In the following we show how to construct the LOs from a subdiagonalization of the LCAO Hamiltonian, 
and we illustrate how to reduce the initial basis set to a subset of LOs through a cut-coupling procedure. 
We also discuss the embedding which will be used for an analysis of the electron occupation. 
All these steps are illustrated in Fig.~\ref{fig:Hstruct}.

\subsection{Subdiagonalization procedure}\label{sec:sub}
 
Let us consider a Hamiltonian $\bb{H}$ and overlap matrix $\bb{S}$ 
obtained from a DFT calculation in the LCAO basis. 
The overlap accounts for the non-orthogonality of the basis functions.

We define the subdiagonalization procedure as follows.
As a first step, we define a set of atoms $S$, which can be a subset $S \subseteq N$ 
of the $N$ atoms in the DFT calculation. 
The choice of $S$ depends on the purpose of the calculation, for instance, 
in order to describe the transport properties of quantum junctions (as we will do in the following), 
the set $S$ can include all atoms of the molecular bridge, with or without anchoring groups. 
In the case of conjugated $\pi$-systems, one could also further restrict 
oneself to the carbon atoms, 
neglecting hydrogen atoms or other functional groups that do not belong to the conjugated system.

We diagonalize each subblock of the selected atoms $i \in S$ 
individually for each atom
and compute the eigenvalues $\{\lambda\}_{i}$ and eigenvectors $\{\ket{\alpha}\}_{i}$. 
Hereafter, we adopt the symbol $\Hm_S$ to refer to this reduced part of the Hamiltonian. 
We now define the block-diagonal projection matrix: 
\begin{equation} \label{eq:Pm}
 \bb{P} = diag( \{\ket{\alpha}\}_{1}, \{\ket{\alpha}\}_{2}, \dots, \{\ket{\alpha}\}_{N} )
\end{equation}
where $\{\ket{\alpha}\}_i$ is a matrix composed of the normalized eigenvectors 
of the atom $i$, when $i \in S$, or an identity matrix otherwise. 
The dimensions of these blocks are determined by the number of basis orbitals 
of the corresponding atomic element. 
The result of performing the Hermitian projection using $\bb{P}$ and its adjoint $\bb{P}^{\dagger}$ 
\begin{equation}\label{eq:Hp}
 \Hp = \Pm^{\dag} \Hm \Pm,
\end{equation}
is to bring the Hamiltonian into a subdiagonal form $\bb{H}^{\prime}$, 
in which the subblocks $i \in N$ on the diagonal, are either diagonal matrices 
containing the eigenvalues $\{\lambda\}_{i}$ or remain unchanged.
The off-diagonal blocks describe the coupling between the atomic orbitals 
in this transformed basis. 
This procedure is illustrated in Fig.~\ref{fig:Hstruct}(b). 
For the overlap matrix $\Sm$, a similar result applies 
except that the diagonal subblocks $i \in S$ are identity matrices. 
In other words, this transforms the LCAO orbitals within $\Hm_S$, 
into a physically interpretable atomic orbitals 
yet deformed by the local chemical environment. 
We refer to these transformed elements as local orbitals (LOs). 
Such LOs are orthogonal within the same atom, but will in general have a finite coupling 
with orbitals on other atoms. 
Indeed, it is this intra-atomic orthogonality that allows for a physical interpretation
of the LOs as atomic orbitals. 

The advantage of looking at the Hamiltonian in this basis representation is that there exist
subsets of LOs that are decoupled from the rest and it is 
sufficient to describe certain physical phenomena where the other subsets can be disregarded. 

This property cannot be inferred in the LCAO basis.
Given the structure of $\Hm$ and $\Sm$ one can define an active space (that will depend on the context) 
and follow two strategies. 
The first one is a cut-coupling procedure, where the LOs outside the active space are neglected. 
The other one is an embedding, where the influence of the other orbitals 
are instead reintroduced via a self-energy matrix.

\subsection{Cut-coupling} \label{sec:cut}
The cut-coupling method can be used as a preprocessing tool to construct 
very accurate and effective Hamiltonians for the study of electronic and transport properties. 
With the subdiagonalization, i.e., the transformation of the LCAO Hamiltonian into a set of well defined LOs, 
it is possible to bring further insight into the results of a DFT calculation 
regarding, e.g., chemical bonding via a selection of a subset of LOs. 
Besides its usefulness in the interpretation of results, the cut-coupling method 
proves especially powerful in reducing the number of basis functions.

\begin{figure}[t!]
    \begin{center}
        \includegraphics[width=0.45\textwidth]{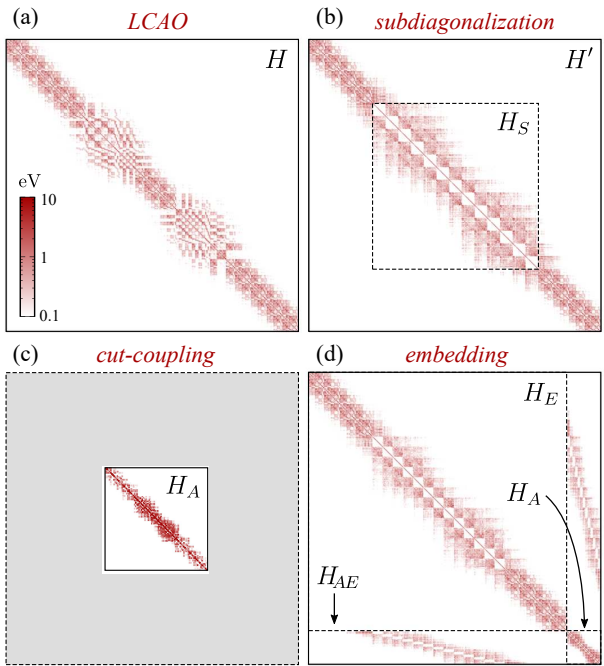}
    \end{center}
    \caption{Schematic illustration of the Hamiltonian transformations discussed in this work. 
    In all cases the value of the element $\Hm_{ij}$ (with $i$ and $j$ composite atom and orbital indices) 
    is represented by its color according to the heatmap in the inset. 
    (a) LCAO Hamiltonian. (b) Hamiltonian in its subdiagonalized form, 
    where each block in the subset $\Hm_S$ was individually subdiagonalized, i.e., transformed into the LO basis. 
    (c) The cut-coupling procedure, where only a few LOs of each subblock of $\Hm_S$ 
    are retained in the active space $\Hm_A$ and the rest (shaded region) are discarded. 
    (d) The embedding procedure, where $\Hm_E$ and the coupling $\Hm_{AE}$ (and its complex conjugate) 
    are used to define an embedding self-energy $\Sigma_A$ for $\Hm_A$.  
    Note that the matrix $\Hm_A$ represented in panels (c) and (d) may differ, 
    depending on the specific application purposes (see Appendix~\ref{app:matrix}). }
    \label{fig:Hstruct}
\end{figure}

If one is interested in describing the physical properties 
close to the Fermi level, such as in the case of electron transport calculations, 
one can select the subset of LOs that include 
the LO eigenvalues closest to the Fermi level, 
which we refer to as the set of relevant LOs. 
In this context, the active space is defined as this set of relevant LOs 
but contains also the AO outside $\Hm_S$.
Practically, this means removing the rows and columns corresponding to all orbitals outside the active space. 
In Fig.~\ref{fig:Hstruct}(b, c) we illustrate the transformation of the Hamiltonian 
with a subdiagonalization procedure followed by the cut-coupling, 
resulting in Hamiltonian $\Hm_A$.
Note that here $\Hm_S$ spans a subset of $\Hp$, 
but in general can also coincide with the whole Hamiltonian 
if elements of the whole system are relevant for the physical properties under consideration. 

The cut-coupling method is generalized for systems with open boundary conditions in Appendix~\ref{app:cut}, 
where the Hamiltonian of the scattering region couples to those of semi-infinite reservoirs 
as in electron transport simulation setups. 


\subsection{Embedding}\label{sec:emb}
The embedding method is an alternative way of making use of LOs, 
for rationalizing and interpreting DFT results. 
Instead of removing a subset of orbitals from the Hamiltonian, 
as in the cut-coupling method, 
we here enclose their effect within a self-energy matrix for the active space. 

The embedding method can be inserted in the framework 
of L\"owdin partitioning approach,~\cite{loewdin_1950,loewdin_1951,loewdin_1962,loewdin_1964} 
generalized to the Green's function formalism, which is a more general method 
in the case of non-orthogonal basis sets, see e.g,, Ref.~\onlinecite{priyadarshy_1996} for a thorough discussion. 
In the context of electron transport, the partitioning technique generally allows to 
separate the scattering region from the semi-infinite leads, enclosing the effects of the latter 
in an embedding self-energy.~\cite{mujica_1994,priyadarshy_1996,datta2005quantum} 
However, here we do something conceptually different.  
We want to partition the Hamiltonian of the scattering region itself, 
separating a subset of degrees of freedom (in this case, a subset of LOs) 
and treat the rest of less relevant degrees of freedom as embedding space. 

In the context of L\"owdin's partitioning technique, we write the Hamiltonian in the block matrix form 
\begin{equation}\label{eq:Hea}
 \Hm = 
     \begin{pmatrix}
     \Hm_E & \Hm_{EA} \\
     \Hm_{AE} & \Hm_A \\
     \end{pmatrix},
\end{equation}
where we have identified the active space and the embedding region 
with the subscripts $A$ and $E$, respectively as illustrated in Fig.~\ref{fig:Hstruct}(d). 

In a single-particle picture this procedure is exact, 
and the partition of the system into active space and embedding region 
can be freely chosen in dependence on the nature of the system and the properties to be evaluated. 
This can be for instance an effective projection of the wavefunction on orbitals with a specific character. 
where the advantage of doing this within a Green's function formalism, 
is to be able to treat also, 
systems with open boundary conditions, or correlated systems, 
where one can include the effect of an additional many-body self-energy on the active space. 

In this case, the Green's function of the active space reads
\begin{equation}\label{eq:GzAemb}
 \Gm_{A}(z) = [z\Sm_A-\Hm_A-\bb{\Sigma}_A(z)]^{-1},
\end{equation}
where $z$ is a complex number and $\bb{\Sigma}_A(z)$ is the embedding self-energy 
with the proper analytic behaviour as defined in Appendix~\ref{app:emb}. 
While this can be done in any basis set, 
we demonstrate that in the LO basis, 
it is possible to find an active space with few LOs 
for which the Green's function $\bb{G}_{A}(z)$ 
accounts for most of the spectral weight around the Fermi level.

\section{Computational Details}\label{sec:gpaw_details}

In the following, we present applications of the method 
to physical systems of general interest. 
In particular we focus on single-molecule junctions, 
but the methodology proposed is generic and can be applied also to 
periodic systems of any structure and chemical composition. 
Unless specified otherwise, the structures were set up 
using the atomic simulation environment (ASE) software package~\cite{larsenJPCM29} 
and DFT calculations were performed with the GPAW package.~\cite{mortensenPRB71,larsenPRB80,enkovaaraJPCM22} 
For converging the electron density, we used an LCAO double-$\zeta$ basis set, 
with a grid spacing of $0.2$~\AA,  
and the Perdew–Burke–Ernzerhof exchange correlation functional.~\cite{perdew_burke_ernzerhof1996}

For the electron transport calculations we used the following setups. 
In one case we consider a broken PA 
junction bridged by a pentacene molecule~\cite{li2011} 
where the scattering region contains six phenylene rings 
on each PA side of the bridge. 
The lead's principal layers are modeled by three unit cells, sampled with a 
$3 \times 1 \times 1$ \textit{k}-point grid along the transport direction. 
For the Au-BDA-Au 
molecular junction, the leads were modeled by a three-layer-thick Au(111) slab 
sampled with a $3\times1\times1$ \textit{k}-point grid along the transport direction. 
The scattering region also includes one Au slab, which is attached to the benzene anchor groups 
via a single atom at the end of a tip. 
In this case, we also performed a geometry optimization, and the atomic positions 
of the BDA were relaxed until the forces on each atom 
were below $0.001$ Hartree bohr$^{-1}$ ($\approx 0.05$ eV/\AA). 
For the bulk calculations of graphene, h-BN, and MoS$_2$ 
we use a $11 \times 11 \times 1$ \textit{k}-point grid to sample the Brillouin zone.

\section{Applications: transport properties}\label{sec:appl_transport}

\subsection{Polyacene organic junctions}\label{sec:PA}
\begin{figure}[b!]
    \begin{center}
        \includegraphics[width=0.5\textwidth]{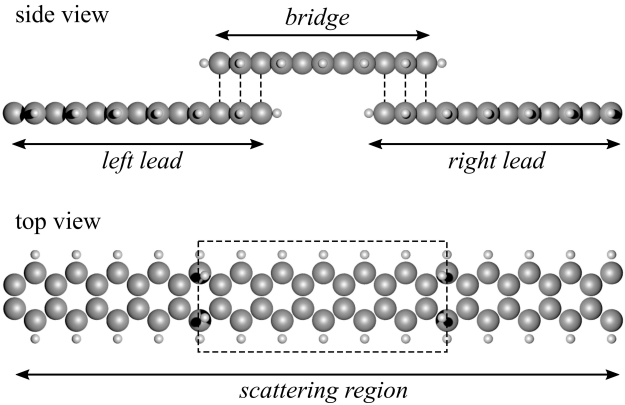}
    \end{center}
    \caption{Side and top view of the junction, consisting of a PA bridge and two PA leads 
    within the scattering region. 
    The bridge is connected to each of the leads via AA stacked phenylene rings. }
    \label{fig:PA_junction}
\end{figure}

We consider a junction with a polyacene (PA) bridge 
connecting PA leads, where the setup is shown in Fig.~\ref{fig:PA_junction}.
We have chosen this structure as representative of a new class of 
all-carbon field effect transistors based on graphene nanoribbons.~\cite{martini2019structure}

\begin{figure}[t!]
    \begin{center}
        \includegraphics[width=0.45\textwidth]{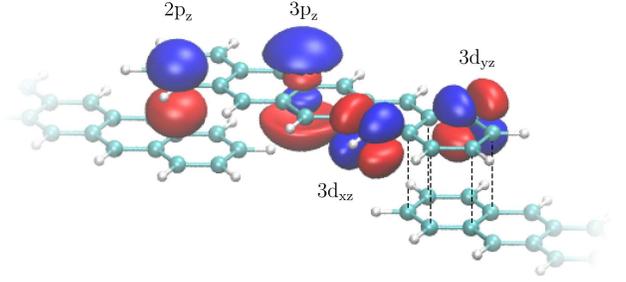}
    \end{center}
    \caption{Schematic visualization of the subset of relevant LOs identified from $\Hm_S$ for the PA junction. 
    Each LO, labelled by its symmetry, is shown for one of the carbon atoms of the bridge PA. 
    The dashed lines highlight the AA stacking between the phenylene rings of the bridge and the lead.}  
    \label{fig:PA_LOs}
\end{figure}

\subsubsection{Identification of a reduced subset of relevant LOs} 

Organic planar molecules are characterized by in-plane $\text{sp}^2$ hybridized orbitals 
accounting for C-C and C-H bonds, while out-of-plane C$\text{2p}_\text{z}$ orbitals 
form $\pi$ molecular orbitals (MOs)  
and enable the delocalization of electrons throughout the molecule. 
Hence, $\pi$ orbitals define the electron transport properties 
of such junctions, 
while the hydrogen atoms are irrelevant and can be neglected in this context. 

With these premises, we subdiagonalize the subblock of the Hamiltonian 
of the scattering region corresponding to each carbom atom. 
As discussed in Sec.~\ref{sec:sub}, we identify the relevant subset of LOs as the ones including 
the LO closest to the Fermi level. 
The definition of the relevant subset becomes evident 
from the analysis of the Hamiltonian matrix, as shown in the Appendix \ref{app:matrix}. 
If we plot the eigenvectors extracted from the corresponding block of the projector, 
we can verify that the most relevant LO on each carbon atom  
closely resembles an atomic $\text{2p}_\text{z}$ orbital. 
A close inspection of the other eigenvectors in the same subset, 
shows that those LOs usually have 
$\text{3p}_\text{z}$, $\text{3d}_\text{xz}$, and $\text{3d}_\text{yz}$ symmetry. 
We refer to the these four LOs for each carbon atom as the relevant subset. 
In Fig.~\ref{fig:PA_LOs} we show all relevant LOs, centered on different atoms for clarity. 

The above observations allow us to define an effective Hamiltonian 
on a reduced basis set for efficient post-processing of DFT calculations. 
In particular, in the following we show that we can obtain a substantial decrease 
of the computational cost for electron transport calculations without sacrificing accuracy.

\subsubsection{Reduced basis set and analysis of electron transport}
The transmission function for phase-coherent transport in a two-terminal device 
with a scattering region connected to bulk reservoirs 
is obtained within the Landauer-B\"{u}ttiker formalism as 
\begin{equation}\label{eq:landauer}
 T(E) = \Tr[\bb{\Gamma}_L \bb{G}^a \bb{\Gamma}_R \bb{G}^r],
\end{equation}
where $\bb{G^{r(a)}}$ is the retarded (advanced) Green's function 
of the scattering region~\footnote{Note that in Appendix~\ref{app:cut} we use the symbol $\Hm_C$ 
for the scattering region.} 
\begin{equation}\label{eq:Gz}
 \bb{G}(z) = [z\Sm - \Hm - \bb{\Sigma}_L(z) -\bb{\Sigma}_R(z)]^{-1}
\end{equation}
with $z=E \pm \imath 0^{+}$ and a self-energy $\bb{\Sigma}_{\alpha}$ 
describing lead $\alpha$. 
See Appendix~\ref{app:cut} for the details. 
In this case the electrodes are periodic PA leads.

\begin{figure}[t!]
    \begin{center}
        \includegraphics[width=0.45\textwidth]{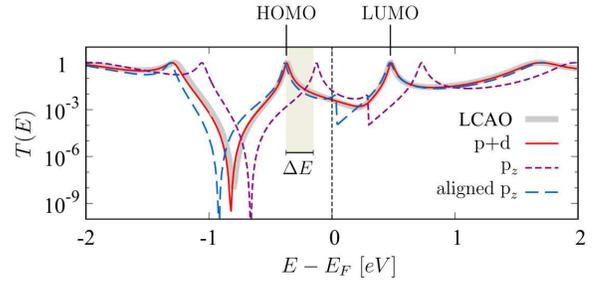}
    \end{center}
    \caption{Transmission function for the PA molecular junctions. 
    The results obtained for the $\text{p}_\text{z}$ and $\text{p$+$d}$ low-energy models 
    are compared to those obtained with the full double-$\zeta$ LCAO set. 
    The transmission within the $\text{p}_\text{z}$ model is also shown 
    with a shift $\Delta E$ in order to align the energy of the HOMO resonance (see text for the details). } 
    \label{fig:trans_PA}
\end{figure}

The computational bottleneck in the evaluation of the transmission is the inversion 
in the definition of the Green's function (\ref{eq:Gz}) for each complex energy $z$. 
Within the recursive Green's function (RGF) technique,~\cite{dmitrytheory} 
the computational cost for inverting the Green's function block by block scales as $mn^3$, 
where $n \times n$ is the typical dimension of one block matrix, and $m$ is the number of such blocks. 
The computational gain to evaluate the transmission in a reduced LOs basis is easily understood, 
because we can now restrict ourselves to a subset of orbitals per atom, thus reducing $n$ for the RGF calculation.~\footnote{Note that the range of inter-atomic interaction, i.e., $m$, remains constant.} 

After the cut-coupling procedure, the Green's function used to evaluate the transmission is given by 
\begin{equation}\label{eq:GzA}
 \bb{G}(z) = [z\Sm_A - \Hm_A - \bb{\Sigma}_L(z) -\bb{\Sigma}_R(z)]^{-1}, 
\end{equation}
where $\bb{\Sigma}_{\alpha}$ are now evaluated for the active space.

We are going to assess the accuracy of the projection by evaluating the transmission function in two cases: 
i) when $\Hm_A$ coincides with the subset of all relevant LOs 
(i.e., those with 
$\text{2p}_\text{z}$, $\text{3p}_\text{z}$, $\text{3d}_\text{xz}$, and $\text{3d}_\text{yz}$ symmetry), 
and ii) when $\Hm_A$ includes only the most relevant $\text{2p}_\text{z}$ LOs. 
We refer to those two approximations as $\text{p$+$d}$ 
and $\text{p}_\text{z}$ models, respectively. 
We remark that the hydrogen atoms are irrelevant for electron transport 
and therefore excluded from $\Hm_A$ in both cases. 
Note that the computational advantage is remarkable 
in both $\text{p$+$d}$ and $\text{p}_\text{z}$ models. 
Considering that in the double-$\zeta$ basis each carbon atom is described by $13$ basis functions,
the computational complexity of the sequential RGF technique is reduced 
to $(4/13)^3 \approx 3\%$ and $(1/13)^3 \approx 0.05\%$ of the original workload, respectively. 
On top of this, neglecting the hydrogen atoms removes $5$ basis functions per hydrogen, 
thus further lowering the typical dimension of one matrix block. 

The results for the transmission function in the PA junction are shown in Fig.~\ref{fig:trans_PA}. 
We observe that the transmission of the $\text{p$+$d}$ model 
is almost indistinguishable from the full one. 
This confirms that the orbitals discarded via the cut-coupling procedure 
do not contribute to the electronic transport within a few eV from the Fermi level. 
Evaluating the transmission within the $\text{p}_\text{z}$ model 
further increases the numerical efficiency of the calculation. 
The corresponding results are qualitatively good, 
but we observe shifts in the position of the transmission resonances. 
Since this systematic error can be in general expected for a $\text{2p}_\text{z}$-only model, 
it is interesting and useful to device a strategy to get rid of the shift. 
The scheme we propose consists in calculating the HOMO of the PA bridge within the full basis set 
and within the $\text{2p}_\text{z}$ model. Thus, we evaluate the energy shift as
$\Delta E=\lambda_{\text{HOMO}}-\lambda^{\text{p}_{z}}_{\text{HOMO}}$, 
and align the transmission $T^{\text{p}_\text{z}}(E-\Delta E)$. 
One can readily verify that, although minor differences are still visible, 
the shifted transmission function represents a reasonable approximation of the full-basis result, 
and in particular, it reproduced with very good accuracy the HOMO-LUMO gap. 

This analysis confirms the \textit{a priori} expectations on the prominent role 
of the $\text{2p}_\text{z}$ orbitals for planar $\text{sp}^2$ systems, 
but it also grants a deeper understanding  of the internal structure of the couplings. 
Moreover, the presented methodology provides a systematic way to improve the approximation 
including a subset of relevant LOs, with an excellent trade-off between efficiency and accuracy.

\subsection{Benzene-diamine (BDA) molecular junction}\label{sec:BDA}
So far, we have demonstrated that, for electron transport calculations for organic planar junctions, 
it is possible to achieve a substantial reduction of the complexity without penalizing the accuracy, 
by identifying the relevant LOs through the subdiagonalization procedure. 
The question arises whether the methodology heavily relies on the presence 
of $\text{sp}^2$ hybridization or whether it can be extended to other systems, 
possibly with some restrictions and {\it caveats}.

\begin{figure}[t!]
    \begin{center}
        \includegraphics[width=0.35\textwidth]{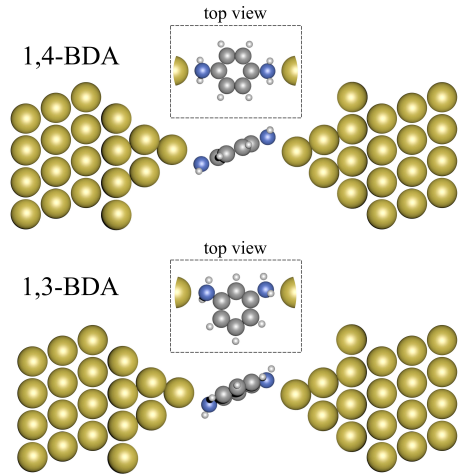}
    \end{center}
    \caption{Side and top view of the 1,4- and 1,3-BDA molecular junctions with Au leads. }
    \label{fig:BDA_junction}
\end{figure}

As a benchmark in this respect, we consider a benzene-diamine (BDA) molecule bridging 
metallic Au(111) leads,~\cite{strangePRB83} which is shown in Fig.~\ref{fig:BDA_junction}
In particular, we focus on two different contact configurations: 
1,3-BDA (meta) and 1,4-BDA (para) which display dramatically different transport properties.  
The aim of this benchmark is two-fold. 
On the one hand, BDA can be considered as the prototypical molecular junction, 
with a relatively simple setup
While the bridge is still organic (and hence we expect the core idea behind our method to hold) 
we expect more complexity in the bonding structure at the interface with the metal, 
to which the carbon atoms are connected via an anchoring group (here NH$_2$) 
we expect changes in the bonding structure, but not a complete overhaul. 
Moreover, in the \emph{meta} connection benzene displays a clear anti-resonance within the HOMO-LUMO gap 
in the electronic transmission function, originating from destructive quantum interference (DQI). 
~\cite{solomonJCP129,markussenNL10,pedersenPRB90,xinJCP146,nozakiJPC121}  
Hence, it is interesting to analyze how the truncated basis set 
is able to describe a specific physical effect, and whether it is possible 
-and to which degree of approximation- to reproduce the DQI features in the transmission. 

Analogously to the case of the PA junction, we subdiagonalize each carbon atom subblock  
of the scattering region. After the cut-coupling procedure  
the active space ($\Hm_A$) includes all LCAO orbitals of N and Au and the relevant LOs of the carbons 
(see Appendix~\ref{app:matrix} for details of the structure of the matrices).
Again, the hydrogen atoms are neglected.

\begin{figure}[t!]
    \begin{center}
        \includegraphics[width=0.5\textwidth]{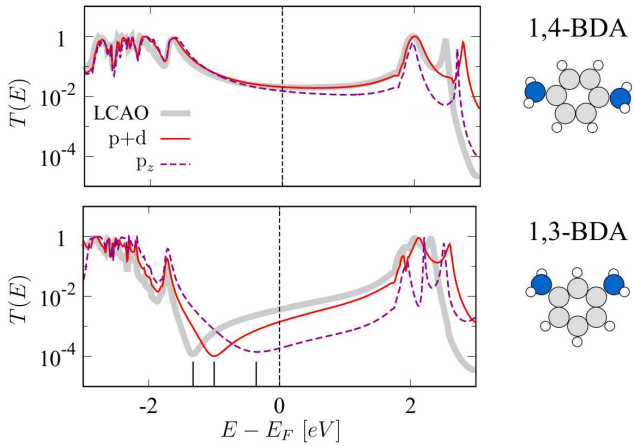}
    \end{center}
    \caption{Transmission function for the 1,4-BDA and 1,3- BDA molecular junctions. 
    The results obtained for the $\text{p}_\text{z}$ and $\text{p$+$d}$ low-energy models 
    are compared to those obtained with the full double-$\zeta$ LCAO set. 
    The position, and the shape of the transmission around the DQI antiresonance 
    are quantitatively affected by the approximation, but overall, 
    the interference patter is reproduced, as well as the position of the HOMO and the LUMO. }
    \label{fig:trans_BDA}
\end{figure}

We evaluate the transmission function $T(E)$ via Eq.~(\ref{eq:landauer}), 
with the Green's function as in Eq.~(\ref{eq:GzA}). 
As before, we consider both the $\text{p$+$d}$ and $\text{p}_\text{z}$ models.~\footnote{Note 
that the $\text{p$+$d}$ model, in this case, includes 
a $\text{3s}$ and a $\text{sp}^3$-like orbital for each external carbon atoms, 
as discussed in Appendix~\ref{app:matrix}, but we do not change the nomenclature for the sake of simplicity. }
In Fig.~\ref{fig:trans_BDA} we compare the reference transmission obtained 
with the full and the reduced basis sets. 
The position of the HOMO and LUMO, and hence the gap, is accurately reproduced by both models. 
In the para configuration, we observe a reduction of the transmission within the HOMO-LUMO gap 
in the $\text{p}_\text{z}$ model, which becomes negligible when all relevant LOs are included. 
In the meta configuration, we observe a clear antiresonance within the HOMO-LUMO gap, 
characteristic feature of DQI. The origin of DQI in meta-connected benzene 
is well established in the literature, and it is ultimately due to a cancellation 
of the coherent superposition of electron waves transmitted across the junction.~\cite{markussenNL10,xinJCP146}
The drastic change in the transmission close to a DQI antiresonance is of interest 
for several applications.~\cite{stadlerNanotech2003,stadlerNanotech2004,stadlerJCP135,valliNL18,valliPRB100}
Thus, the possibility of reproducing such features of the transmission function 
with an effective model is an important application   
of the methodology proposed in this work. 
The position of the antiresonance and the shape of the transmission function around it  
seems to strongly depend on the basis set used for the calculation. 
Specifically, within the  $\text{p}_\text{z}$ model, the transmission displays a broad suppression, 
centered close to the Fermi level, with a relatively symmetric shape. 
In the $\text{p$+$d}$ model, the antiresonance becomes sharper and shifts closer to the HOMO, 
and resembles very closely the shape of the transmission function with the full basis set. 
Despite the differences described above, for all basis sets 
there is a clear suppression of the transmission at the Fermi level 
for the meta configuration compared to the para configuration, 
which is eventually what is observed from the analysis of conductance histograms 
in standard experimental setups. 

\section{Applications: electronic properties}\label{sec:appl_electr}

We have shown in the previous section that the LO basis set is useful for an efficient evaluation 
of electron transport properties in single-molecule junctions. 
In addition to the computational advantage, the LOs constitute a powerful tool to analyse and interpret conductance spectra. 
Here, we motivate applications of LOs 
as a basis suitable to postprocess electronic structure calculations 
and to construct effective \textit{ab initio} tight-binding Hamiltonians 
which could potentially be used for many-body methods beyond DFT. 
Using BDA as a benchmark system, we show again how 
the cut-coupling and embedding procedures can be used to make 
a complete characterization of the electronic properties of molecular junctions 
from an analysis of the frontier MOs. 


\subsection{Effective $\text{p}_\text{z}$ Hamiltonians of BDA junction.}\label{sec:embed_BDA}



Using the BDA example, we make an analysis of the MOs associated to the central molecule 
in terms of relevant LOs. 
This analysis is tackled from two perspectives. On one side, 
we expand the frontier MOs in terms of a combination of LOs with $\text{p}_\text{z}$ character, for each C atom and for each N atoms, and show that this minimal basis 
set yields an electronic distribution in close agreement with the corresponding LCAO distribution. 
Furthermore, we compare the result with the MOs constructed from the original set of $\text{p}_\text{z}$ AOs from the LCAO basis set and we shown that the latter is less suited 
for constructing effective model Hamiltonians, as these AOs, contrarily to LOs, lack the information 
corresponding to the environment. On the other side, we decompose the density of states (DOS) into the  projected DOS 
associated with the $\text{p}_\text{z}$ LOs and demonstrate that the latter account for most of the spectral weight around the Fermi level.

\begin{figure}[t!]
    \begin{center}
        \includegraphics[width=0.5\textwidth]{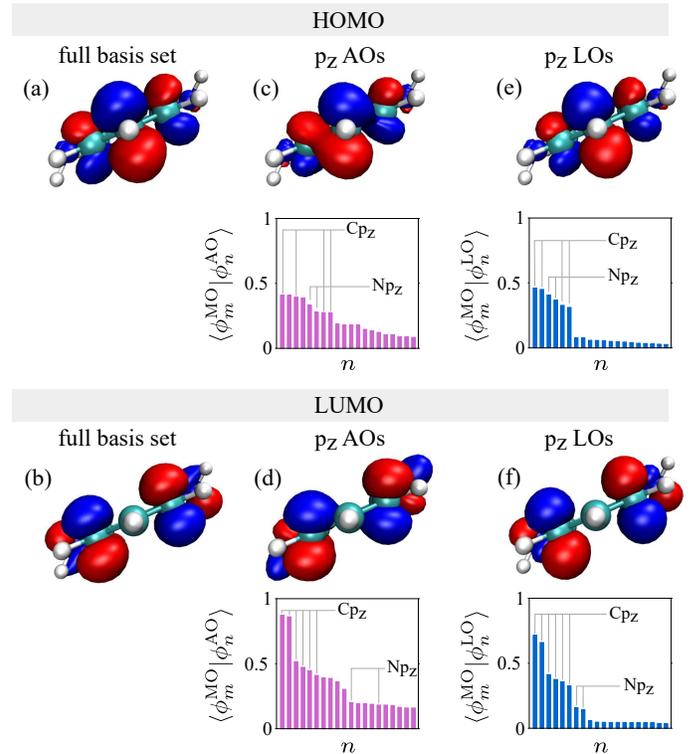}
    \end{center}
    \caption{Frontier MOs of the 1,3-BDA molecule (a, b) 
    and their projection onto carbon and nitrogen $\text{p}_\text{z}$ orbitals 
    in the AO (c,d ) and in the LO (e,f) basis.
    For each projection, we also show a few sorted expansion coefficients 
    $\langle \phi_m^{\text{MO}}|\phi_n \rangle$ in the corresponding (AOs or LOs) bases, 
    with those corresponding to the $\text{p}_\text{z}$ orbitals explicitly labelled. }
    \label{fig:meta_BDA_MOs}
\end{figure}

\begin{figure}[h!]
    \begin{center}
        \includegraphics[width=0.5\textwidth]{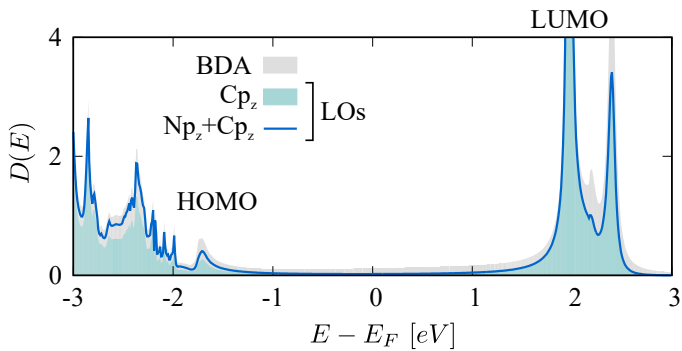}
    \end{center}
    \caption{Projected DOS $D_i(E)$ the 1,3-BDA molecular junction, onto the whole molecule (grey shaded area), 
    onto the $\text{p}_\text{z}$ LOs of carbon (cyan shaded area) or carbon and nitrogen (solid line). 
    In the LOs basis, most of the spectral weight is projected of the molecule 
    onto the orbitals with $\text{p}_\text{z}$ character. }
    \label{fig:meta_BDA_gs}
\end{figure}

For the calculation of the MOs, we diagonalize the BDA Hamiltonian, 
i.e., the Hamiltonian sub-block that includes the benzene molecule and the amino groups. 
In Fig.~\ref{fig:meta_BDA_MOs}(a,b) we show the frontier MOs of the 1,3-BDA molecule. 
Each MO is determined by an eigenvector whose elements represent 
the coefficients $\langle \phi_m^{\text{MO}}|\phi_n^{\text{AO}} \rangle$ 
of its expansion in terms of the AO basis. 
By selecting only the coefficient corresponding to the $\text{p}_\text{z}$ AO, 
we can display the projection of the MO onto those orbitals. 
Since the frontier MOs have a significant weight on the nitrogen atoms, 
we include both C$\text{p}_\text{z}$ and N$\text{p}_\text{z}$ in the projection. 
This is shown in Fig.~\ref{fig:meta_BDA_MOs}(c,d). 
One can already notice significant differences (also in terms of symmetries) 
between each frontier MO and its AOs projections. 
In particular, the projected MO is skewed towards the $z$ axis of the reference frame 
of the scattering region, where the $\text{p}_\text{z}$ AOs are oriented. 
This can be understood by looking at the sorted distribution of the expansion coefficients.  
For both frontier MOs, the coefficients corresponding to the $\text{p}_\text{z}$ orbitals 
are among those with the highest values in the histograms,   
but the distribution have a long "tail" with sizable contributions 
from many orbitals with different symmetries, which also include orbitals centered on hydrogen atoms. 
As a consequence, the projection onto the $\text{p}_\text{z}$ AOs yields 
a poor approximation of the original frontier MO. 

Analogously, after the subdiagonalization of all atoms of the BDA Hamiltonian, 
we can consider the projection onto the $\text{p}_\text{z}$-like LOs of the carbon and nitrogen atoms,  
which is shown in Fig.~\ref{fig:meta_BDA_MOs}(e,f). 
It is evident that the projection onto the LO basis yields an electronic distribution 
that resemble more closely the original MOs than the one obtained in the AO basis. 
As anticipated, one of the reasons is that the $\text{p}_\text{z}$-like LOs 
are modified by the local chemical environment and are therefore oriented along 
the axis perpendicular to the benzene plane, 
which does not coincide with the $z$ axis (cfr. Fig.~\ref{fig:BDA_junction}). 
This is confirmed by the distribution of the coefficients 
$\langle \phi_m^{\text{MO}}|\phi_n^{\text{LO}} \rangle$, 
which are now the elements of the transformation that brings the Hamiltonian 
from the subdiagonalized to its diagonal form. 
The highest coefficients are again those of the $\text{p}_\text{z}$ LOs, 
but this time the mixing with orbitals of other symmetries is weak, 
resulting in an overall better projection of the frontier MOs onto orbitals 
with exclusively $\text{p}_\text{z}$ character.

A complementary analysis consists in looking at the projected DOS 
\begin{equation}
 D(E) = \frac{1}{2\pi} \Tr [\Am_{A} \Sm_A],  
\end{equation}
where $\Am_A=\imath(\Gm_A-\Gm_A^{\dagger})$ is expressed through the 
Green's function of the active space $\Gm_A(z)$ of Eq.~(\ref{eq:GzAemb}),
evaluated at $z=E+\imath 0^{+}$. 
In contrast to the MO analysis, where we only consider the Hamiltonian of the BDA, 
this has the advantage that all orbitals outside the active space 
are taken into account through an embedding self-energy. 
In particular, we evaluate $D(E)$ for three cases: i) we consider the whole BDA molecule as the active space, 
while the embedding includes the effect of the Au orbitals in the scattering region and the leads, 
while in the other two cases the active space consists only of LOs with 
ii) C$\text{p}_\text{z}$ character and iii) C$\text{p}_\text{z}$ and N$\text{p}_\text{z}$ character, 
and the rest of the molecule belongs to the embedding together with the Au orbitals. 
It is known that care must be taken when projecting 
with a non-orthogonal basis set.~\cite{sorianoPRB90,jacobJPCM27} 
Here, in each case, we orthogonalize the active and embedding subspaces 
as discussed in Appendix~\ref{app:emb}. 

The results are shown in Fig.~\ref{fig:meta_BDA_gs} for the 1,3-BDA junction. 
Most of the spectral weight projected onto the BDA molecule is accounted for by LOs 
with $\text{p}_\text{z}$ character.  
For the LUMO resonance the C$\text{p}_\text{z}$ is dominant, 
while around the HOMO also the contribution of N$\text{p}_\text{z}$ is not negligible. 
This mirrors the conclusions we drew from the analysis of the distribution 
of the expansion coefficients of the frontier MOs. 

The above analysis demonstrates that the LOs represent a better basis set than the original AOs 
for post-processing, and are particularly useful to build an effective $\text{p}_\text{z}$ model for BDA. 
Moreover, reduced basis sets of LOs can be used, e.g., for many-body calculations beyond DFT, 
which would be numerically challenging  (and in general prohibitive for complex systems) 
to perform in a full LCAO basis set.

\begin{figure*}
    \begin{center}
        \includegraphics[width=1.0\textwidth]{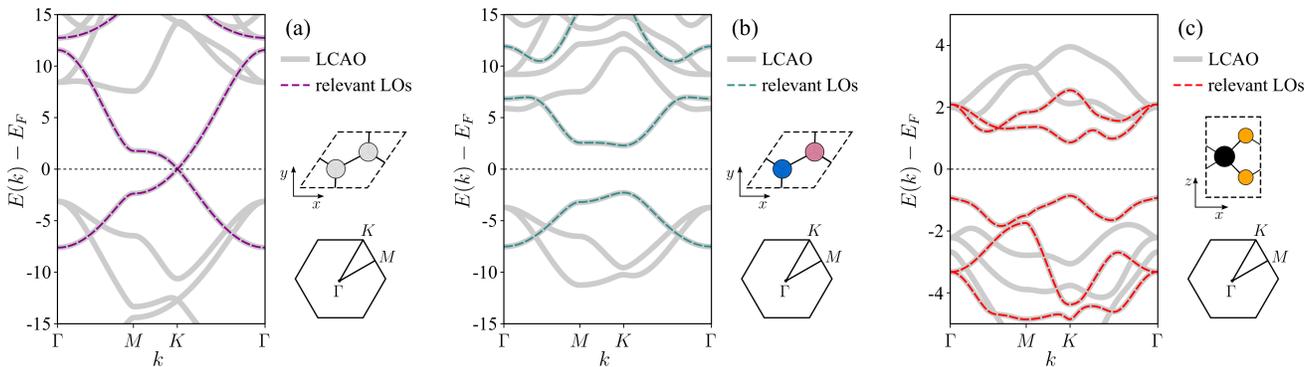}
    \end{center}
    \caption{Bandstructure, unit cell, and Brillouin zone 
    of (a) graphene, (b) h-BN, and (c) MoS$_2$ monolayers. 
    The results obtained with the reduced basis set of the relevant LOs (dashed lines) 
    reproduce the low-energy bandstructure from the full double-$\zeta$ LCAO 
    basis set calculation (grey solid lines). }
    \label{fig:bands}
\end{figure*}

\section{Extension to periodic systems} \label{sec:bands}
So far, we have shown that the subdiagonalization of the Hamiltonian allows to identify 
a reduced basis set of relevant LOs, which allows to perform post-processing and analysis 
of DFT calculations more efficiently without sacrificing the accuracy. 
In particular, we demonstrated that the subset of relevant LOs 
is sufficient to describe electron transport around the Fermi level in molecular junctions. 
By inspecting the structure of the Hamiltonian matrices,
we realized that the relevant LOs have a sizable coupling only among themselves. 
Indeed, it is this closure relation that determines the accuracy of the reduced models. 
A natural question arises whether similar patterns are found in different materials other than organic compounds.
Furthermore, up until now we have focused on finite-size systems in combination 
with open boundary conditions. 
In this section, we extend the methodology to the study of the electronic properties 
of periodic systems and provide a few notable examples. 

\subsection{General strategy for subdiagonalization and cut-coupling}

Let us introduce the notation $\Hm(\Rv)$, where $\Rv$ is a real-space lattice vector. 
The matrix element $H_{mn}(\Rv=\Ov)=H_{mn}(\Ov)$ describes the coupling 
between atomic orbitals $m$ and $n$ within the first unit cell. 
In a material with more than one atom per unit cell, this term can describe 
both intra-atomic and iter-atomic interactions. 
The matrix element $H_{mn}(\Rv\neq\Ov)$ denote the interaction between 
atomic orbitals $m$ and $n$ with one located in the first unit cell 
and one in the periodic repetitions, respectively.  

In analogy with the case for open boundary calculations, we construct the projector $\Pm$ that subdiagonalizes each atom in the first unit cell: 
\begin{equation}\label{eq:Hp_Ov}
\begin{aligned}
 \Hp(\Ov) = \Pm^{\dag} \Hm(\Ov) \Pm.
\end{aligned}
\end{equation}
Hence, in periodic systems $\Hm_S$ coincides with $\Hm(\Ov)$.
In order to properly account for this transformation, we additionally 
need to rotate the couplings $H_{mn}(\Rv)$
\begin{equation}\label{eq:Hp_Rv}
\begin{aligned}
 \Hp(\Rv) = \Pm^{\dag} \Hm(\Rv) \Pm \quad \forall \Rv,
\end{aligned}
\end{equation}
with the projector defined in Eq.~(\ref{eq:Hp_Ov}). 
The set of transformations in Eqs.~(\ref{eq:Hp_Ov}) and (\ref{eq:Hp_Rv}) 
together with the corresponding ones for the overlap matrix $\Sm(\Rv)$ 
constitute a unitary transformation between LCAO and LO basis for a periodic system. 

In order to define the active space, we need to inspect the structure of the couplings 
between atoms within the first unit cell, and between different cells.
If it is possible to identify a subset of relevant LOs weakly coupled with the rest of the basis, 
we can perform the cut-coupling procedure for both $\Hm(\Ov)$ and $\Hm(\Rv)$. 
The general scheme to achieve this is as follows. 
We start by identifying from the diagonal elements of $\Hm(\Ov)$ the LO with energy 
closest to the Fermi level, which we denote with the index $\ell$. 
Without loss of generality, let us suppose that this $\text{LO}$ belongs to atomic site $A$. 
Next, we need to consider the following two types of couplings: 
i) The first type are the intra-atomic couplings between $\ell$ within the first unit cell 
and the other LOs centered at one of the periodic repetitions of the same atomic site. 
These terms are given by the off-diagonal matrix elements $H_{\ell a}(\Rv\neq\Ov)$, 
where $a$ spans all LOs of site $A$. 
We note that the LOs within a given atom 
are by construction orthogonal. 
ii) The second type are the inter-atomic couplings between $\ell$ and the other LOs 
centered at other atoms within the first unit cell. 
These terms are given by the off-diagonal matrix elements $H_{\ell b}(\Ov)$, 
where $b$ spans all LOs of sites $B\neq A$.
Note that these terms are absent in single-atom unit cells.
For both types we select those LOs that have a sizable coupling with $\ell$.
Finally, we construct the relevant set of LOs by uniting the two sets defined above. 
Inter-atomic couplings between different atomic sites in different unit cells 
follow the same pattern as those of type ii) for any value of $\Rv$. 

The Hamiltonian in the reciprocal space can be obtained by
\begin{equation}
  H(\kv) = \sum_{\Rv} H(\Rv) e^{\imath \kv \Rv} 
\end{equation} 
where $\kv$ is a reciprocal lattice vector in the first Brillouin zone. 
The dispersion relation for the $i$-th band at any point $\kv$ is given 
by the eigenvalue $E_i(\kv)$ of the Hamiltonian.

\subsection{Bandstructure of graphene, h-BN, and MoS$_2$ monolayer}
We compute the bandstructure $E(\kv)$ 
of graphene, hexagonal boron nitrade (h-BN) and molybdenum disufide (MoS$_2$) 
The application to graphene aims at demonstrating the subdiagonalization procedure for periodic organic systems. Moreover, we assess its validity to study the electronic structure.  
The case of h-BN is a generalization to a 2D system with a heteroatomic unit cell. 
and MoS$_2$ is a study case for a quasi-2D transition-metal dichalcogenide monolayer. 
In Fig.~\ref{fig:bands} we show the bandstructure obtained with the full LCAO basis 
and in the relevant LO basis for all these systems. 
Since all systems considered here have a hexagonal Brillouin zone, 
the bandstructure is shown along a path through the high-symmetry points $\Gamma-M-K-\Gamma$. 

For graphene the set of relevant LOs of each carbon atom in the unit cell have 
$\text{2p}_\text{z}$, $\text{3p}_\text{z}$, $\text{3d}_\text{xz}$, and $\text{3d}_\text{yz}$ symmetry, 
in complete analogy to the organic junctions discussed above. 
The LCAO graphene bandstructure close to the Fermi level, 
including the position and the degeneracy of the Dirac point. 
is accurately reproduced by the subset of relevant LOs, see Fig.~\ref{fig:bands}(a), as expected. 
 
For h-BN we obtain $8$ relevant LOs ($4$ centered at the B site and the other $4$ at the N site). 
Also in this case, the bands in the reduced LO basis are indistinguishable 
from the corresponding LCAO ones, see Fig.~\ref{fig:bands}(b), 
and accurately reproduce the direct bandgap at the $K$ point, where the valence and conduction bands 
have predominantly N$\text{2p}_\text{z}$ and B$\text{2p}_\text{z}$ character, respectively. 

Finally we turn to the MoS$_2$. 
Here, the reduced basis set consists of more LOs with respect to the previous cases,  
i.e., $22$ LOs (out of the original $55$ LCAOs), 
partly due to the non-planarity of the dichalcogenide structure. 
However, also in this case, the bandstructure close to the Fermi level 
and the direct band gap at $K$ are accurately reproduced, see Fig.~\ref{fig:bands}(c). 

In all cases considered here, it was always possible to identify a subset of LOs, 
which is weakly coupled with the rest of the basis set 
and describes very accurately the electronic bandstructure close to the Fermi level. 
This procedure is suitable to obtain reliable approximations of the bandstructure 
with a substantially reduced computational cost, useful, 
e.g., for transport calculations in the framework of 
a top-of-the barrier model and effective-mass approximation,~\cite{rahman2003theory} 
or electronic calculations, such as obtaining few-orbitals tight-binding models with DFT parameters, 
or interpolation schemes, by which quantities computed on a relatively coarse k-space mesh 
can be used to interpolate faithfully onto an arbitrarily fine k-space mesh at relatively low cost.

\section{Summary and Outlook} \label{sec:outlook}

We proposed a method to identify a subset of orbitals, 
which yield an accurate description of the electronic structure close to the Fermi level. 
By essentially removing the redundancy of LCAO basis sets, 
it allows for an efficient calculation of transport and electronic properties 
which can be derived from a DFT simulations, without sacrificing accuracy. 
The method can be applied to both molecular junctions and periodic systems,  
and it is based on the subdiagonalization of the LCAO Hamiltonian on each atom 
of the scattering region or periodic unit cell. 
This corresponds to a transformation to a basis of LOs. 
We observe the emergence of clear coupling patterns between the LOs, 
which are not apparent in the original LCAO basis. 
In particular, we can always identify a subset of LOs (denoted relevant LOs) 
which to a first approximation is decoupled from the rest of the basis set. 
Performing calculations restricting the active space to the subset of relevant LOs, 
we can achieve a substantial reduction of the computational costs 
while retaining an accuracy comparable with calculation performed with the full basis set. 

We illustrate the potential of the method focusing on a few selected applications. 
Namely, we compute the transmission function of prototypical junctions, 
such as a PA molecule bridging PA leads and a BDA molecule bridging Au(111) leads. 
To demonstrate the possibility of applications to periodic systems, 
we compute the bandstructure of graphene, h-BN, and MoS$_2$ monolayers. 

Finally, we stress that there is the potential for 
several other interesting applications for the proposed methodology. 
For instance, it could be helpful to understand the effects of functionalization of molecules 
or adsorption of atoms or molecules on surfaces, 
by performing analysis in both the LOs and the MOs bases. 
Moreover, the combination of the subdiagonalization and embedding procedures  
could be employed to define effective tight-binding models on a reduced basis set. 
The latter is appealing, e.g., for tight-binding parametrizations 
of real materials~\cite{jacobJCP134,calogeroNS11} 
and for methods suitable to address strong electronic correlations, 
such as GW and the dynamical mean-field theory~\cite{georgesRMP68} (DMFT) 
and its real-space extension~\cite{potthoffPRB60} 
aiming at the description of inhomogeneous~\cite{snoekNJP10,amaricciPRA89,baumannPRA101} 
and nanoscopic systems.~\cite{valliPRL104,jacobPRB82,valliPRB86,dasPRL107,valliPRB91,valliPRB92,jacobJPCM27,valliPRB94,schuelerEPJBST226,valliNL18,pudleinerPRB99,kropfPRB100,valliPRB100} 
In particular, due to their spatial localization, the LOs represent a possible alternative 
to Wannier orbitals~\cite{marzariRMP84} or natural orbitals~\cite{simPRB100} 
to define local Coulomb interaction parameters 
in the framework of a DFT+DMFT and GW+DMFT approaches (for recent reviews of these topics see, 
e.g., Refs.~\onlinecite{kotliarRMP78,heldAP56,schuelerEPJBST226,tomczakEJPBST226}). 
We believe this to be a particularly promising route for tackling 
electronic correlation effects in a wide class of nanostructures, ranging from graphene nanoribbons 
to organo-metallic complexes, such as transition metal porphyrins and phthalocyanines.

\begin{acknowledgments}
We thank F.~Libisch, M.~Luisier, L.~Mennel, J.~M.~Tomczak for insightful discussions. 
A.V. and R.S. acknowledge financial support from the Austrian Science Fund (FWF) through project P~31631. 
G.G. and D.P. acknowledge financial support from NCCR MARVEL funded 
by the Swiss National Science Foundation (51NF40-182892). 
Some numerical calculations have been performed on the Vienna Scientific Cluster (Project No. 71279) 
and the internal cluster in the ETHZ Integrated Systems Laboratory. 
\end{acknowledgments}

\section*{DATA AVAILABILITY}
The data that support the findings of this study are available from the corresponding author upon reasonable request.

\appendix

\section{Cut-coupling} \label{app:cut}
In electron transport simulation schemes, a finite-size scattering region, or central region, 
is treated with open boundary conditions and is coupled to two (or more) charge reservoirs (leads) 
which are modeled by a periodic repetition of a bulk unit cell. 
Focusing on a two-terminal setup, the Hamiltonian expanding the coupled system 
can be partitioned into matrix blocks corresponding to the respective regions in real space: 
\begin{equation}\label{eq:H_LCR}
 \Hm =  
     \begin{pmatrix}
     \Hm_L & \Hm_{LC} & 0 \\
     \Hm_{CL} & \Hm_{C} & \Hm_{CR} \\
     0 & \Hm_{RC} & \Hm_{R} \\
     \end{pmatrix},
\end{equation}
where $L$, $R$ and $C$ denote the left lead, right lead and central region, respectively. 
The central region is chosen to be wide enough that the wavefunctions of the leads do not overlap, 
so that without loss of generality we can assume $\Hm_{LR}=\Hm_{RL}=0$. 
We construct the projector $\Pm$ in Eq.~(\ref{eq:Pm}) such that 
\begin{equation}\label{eq:Hp_C}
 \Hp_C = \Pm^{\dagger} \Hm_C \Pm 
\end{equation}
is the subdiagonalized Hamiltonian for the scattering region. 
The coupled system in this transformed basis is given by: 
\begin{equation}\label{eq:Hp_LCR}
 \Hp =  
     \begin{pmatrix}
     \Hm_L & \Hp_{LC} & 0 \\
     \Hp_{CL} & \Hp_{C} & \Hp_{CR} \\
     0 & \Hp_{RC} & \Hm_{R} \\
     \end{pmatrix}.
\end{equation}
where ($\alpha \in {L,R}$)
\begin{equation}\label{eq:Hp_LR}
\begin{aligned}
    \Hp_{\alpha C} &= \Hm_{\alpha C} \Pm, \\
    \Hp_{C \alpha} &= \Pm^{\dagger} \Hm_{C \alpha}.
\end{aligned}
\end{equation}
Note that the Hamiltonian matrices $\Hm_L$ and $\Hm_R$ in Eq.~(\ref{eq:Hp_LCR}) 
are unaffected by the projection, which means that the associated surface Green's functions 
remain exactly the same as those computed in the original basis. 
Similar transformations are required for the overlap matrix.

We remove the rows and the columns corresponding to all LOs outside the active space 
in $\Hp_C$ (i.e., within the scattering region) 
as well as in $\Hp_{\alpha C}$ and $\Hp_{C \alpha}$ (i.e., to the leads). 
This identifies the Hamiltonian of the active space $\Hm_A$ 
which can be used for efficient transport calculations 
by defining the Green's function of Eq.~(\ref{eq:GzA}) as 
\begin{equation}
 \bb{G}(z) = [z\Sm_A - \Hm_A - \bb{\Sigma}_L(z) -\bb{\Sigma}_R(z)]^{-1}, \tag{\ref{eq:GzA}}
\end{equation}
where $z = E + \imath 0^{+}$ and $\bb{\Sigma}_{\alpha}(z)$ are the self-energies
of the left and right leads. 


\section{Embedding}\label{app:emb} 


For the embedding we regroup and reorder $\Hp$ (i.e., the subdiagonalized Hamiltonian) 
such that the structure of the Hamiltonian is
\begin{equation} 
 \Ht = 
     \begin{pmatrix}
     \Hm_E & \Hm_{EA} \\
     \Hm_{AE} & \Hm_A \\
     \end{pmatrix}, \tag{\ref{eq:Hea}}
\end{equation}
and similarly 
\begin{equation}\label{eq:SeaP}
 \St = 
     \begin{pmatrix}
     \Sm_E & \Sm_{EA} \\
     \Sm_{AE} & \Sm_A \\
     \end{pmatrix},
\end{equation}
for the overlap matrix. Here, we have identified the active space and the embedding region 
with the subscripts $A$ and $E$, respectively. 
We describe the effect of the surroundings $\Hm_E$ on the active space 
by a self-energy matrix $\bb{\Sigma}_A(z)$ as~\footnote{For the details, see, e.g., the book of Datta, chapter 8.4, Eq.~(8.4.7).~\cite{datta2005quantum}}
\begin{equation} \label{eq:GzAemb_badhe}
 \bb{G}_{A}(z) = [z\Sm_A-\Hm_A-\bb{\Sigma}_A(z)]^{-1}. 
\end{equation}
Particular care must be taken when extending the formalism to non-orthogonal basis functions. 
It has already been argued on a more formal ground, that the choice of the projector $\Pm$ in Eq.~(\ref{eq:Pm}) 
yields a tensorial inconsistent density matrix for the active space.~\cite{o2012subspace,sorianoPRB90,jacobJPCM27} 
The problem arises due to non vanishing overlap $\Sm_{EA}(\Sm_{AE})$ between the embedding region 
and the active space. 
A possible solution is to orthogonalize the two regions,~\cite{Thygesen_2006,Kwok_2013,Droghetti_2017} 
i.e., we seek for a basis transformation 
upon which $\Sm_{EA} = \Sm_{AE} = 0$, but leaves the basis in the active space unchanged, 
so that the atomic character of the LOs is preserved:
\begin{equation} \label{eq:H2x2}
 \Hb = \Um^{\dag} \Ht \Um = \begin{pmatrix}
     \Hb_E & \Hb_{EA} \\
     \Hb_{AE} & \Hm_{A} \\
 \end{pmatrix},
\end{equation}
\begin{equation} \label{eq:S2x2}
 \Sb = \Um^{\dag} \Sm \Um = \begin{pmatrix}
     \Sb_E & \boldsymbol{0} \\
     \boldsymbol{0} & \Sm_{A} \\
     \end{pmatrix}.
\end{equation}
This is obtained with the following transformation:
\begin{equation} \label{eq:U2x2}
 \Um = \begin{pmatrix}
     \bb{1}_E & \bb{0} \\
     -\Sm^{-1}_A \Sm_{AE} & \bb{1}_A \\
       \end{pmatrix}.
\end{equation}
Here, we have introduced the identity matrix $\bb{1}_m$ of dimensions $m \times m$. The transformed overlap and Hamiltonian matrices are
\begin{equation}
 \begin{aligned}
     \Sb_E = \ & \Sm_E - \Sm_{EA} \Sm_A^{-1} \Sm_{AE}, \\[3pt]
     \Hb_E = \ & \Hm_E + \Sm_{EA} \Sm_A^{-1} \Hm_{A} \Sm_A^{-1} \Sm_{AE}  \\
     & - \Hm_{EA} \Sm_A^{-1} \Sm_{AE} - \Sm_{EA} \Sm_A^{-1} \Hm_{AE}, \\[3pt]
     \Hb_{EA} = \ & \Hm_{EA} - \Sm_{EA} \Sm_A^{-1} \Hm_A,
 \end{aligned}
\end{equation}
where we used a bar on top of the symbols to denote the matrices in the transformed basis. 
The embedding self-energy can then be written as
\begin{equation} \label{eq:Sigemb}
 \Sigb_A(z) = \Hb_{AE}\gb_E(z)\Hb_{EA},
\end{equation}
where
\begin{equation}
 \gb_E(z) = (z\Sb_E-\Hb_E)^{-1}
\end{equation}
is the bare Green's function for the isolated embedding region. 
Under this transformation, in the limit as $z$ goes to infinity, $\gb_E(z) \propto 1/z$ 
and the embedding self-energy displays the correct physical decay. 
It is easy to see that, without orthogonalization, the coupling matrices in Eq.~(\ref{eq:Sigemb}) 
are replaced by $\Hb_{AE} \rightarrow z\Sm_{AE}-\Hm_{AE}$ and $\Hb_{EA} \rightarrow z\Sm_{EA}-\Hm_{EA}$, 
thus spoiling the high-energy behavior. 

The Green's function of the active space in given by  
\begin{equation} \label{eq:GzAemb_goodhe}
 \Gb_{A}(z) = [z\Sm_A-\Hm_A-\Sigb_A(z)]^{-1} 
\end{equation} 
which, in contrast to Eq.~(\ref{eq:GzAemb_badhe}), displays the proper high-energy behavior. 
In Eq.~(\ref{eq:GzAemb}) and throughout the text we use the notation $\Gm_A(z)$, 
but compute the corresponding quantities with Eq.~(\ref{eq:GzAemb_goodhe}) instead.

Finally, if the embedding region is coupled to other systems, such as the electrodes 
in an electron transport calculation setup, the Green's function of the embedding is given by
\begin{equation} \label{eq:gE}
 \gb_E(z) = [z\Sb_E-\Hb_E-\sum_{\alpha}\bb{\Sigma}_{\alpha}(z)]^{-1},
\end{equation}
where the $\bb{\Sigma}_{\alpha}(z)$ is the lead's self-energy, 
with $\alpha = L, R$ in the case of a two-terminal setup 
\begin{equation}
  \bb{\Sigma}_{\alpha}(z) = [z\Sm_{\alpha E}-\Hm_{\alpha E}]\bb{g}_{\alpha}(z) [z\Sm_{\alpha E}^{\dag}-\Hm_{\alpha E}^{\dag}].
\end{equation}
Here, $\Sm_{\alpha E}$ and $\Hm_{\alpha E}$ are the off-diagonal matrix blocks that describe the overlap and coupling to the lead $\alpha$ and $\bb{g}_{\alpha}(z)$ is the corresponding surface Green's function. 
Note that the leads' self-energy is not changed by the transformation. 
This is readily seen by generalizing Eqs.~(\ref{eq:H2x2}-\ref{eq:U2x2}) to explicitly include 
a sublock for a lead $\alpha$: 
\begin{equation}
 \Hm = 
     \begin{pmatrix}
     \Hm_{\alpha}   & \Hm_{\alpha E} & \bb{0} \\
     \Hm_{E\alpha} & \Hm_E              & \Hm_{EA} \\
     \bb{0} & \Hm_{AE}         & \Hm_A \\
     \end{pmatrix},
\end{equation} 
\begin{equation}
 \Sm = 
     \begin{pmatrix}
     \Sm_{\alpha}   & \Sm_{\alpha E} & \bb{0} \\
     \Sm_{E\alpha} & \Sm_E              & \Sm_{EA} \\
     \bb{0}              & \Sm_{AE}         & \Sm_A \\
     \end{pmatrix}.
\end{equation} 
We note that both conditions $\Hm_{\alpha A}=0$ and $\Sm_{\alpha A}=0$ 
are ensured by including part of the lead at the boundaries of the scattering region. 
The corresponding transformation then reads
\begin{equation}\label{eq:U3x3}
 \Um = 
     \begin{pmatrix}
     \bb{1}_{\alpha} & \bb{0}      & \bb{0} \\
     \bb{0}          & \bb{1}_E    & \bb{0} \\
     \bb{0}          & \Sm_{A}^{-1}\Sm_{AE} & \bb{1}_A \\
     \end{pmatrix}.
\end{equation}
Hence the transformation $\Hb = \Um^{\dagger} \Hm \Um$ and $\Sb = \Um^{\dagger} \Sm \Um$ 
leave the $\Hm_{\alpha E}$ and $\Sm_{\alpha E}$ subblocks unchanged. 

For the sake of completeness, we note that the transformation $\Um$ 
in Eqs.~(\ref{eq:U2x2}) and (\ref{eq:U3x3})
is norm-conserving and preserves the whole spectrum, 
so that any calculated quantity remains unaffected.~\cite{Thygesen_2006}

\section{Structure of the subdiagonalized Hamiltonian} \label{app:matrix}
Here we show explicitly the structure of the subdiagonalized Hamiltonian 
for the PA and BDA junctions analyzed in this work.

\subsection{PA: Subdiagonalization for all-carbon junctions}
\begin{figure}[t!]
    \begin{center}
        \includegraphics[width=0.4\textwidth]{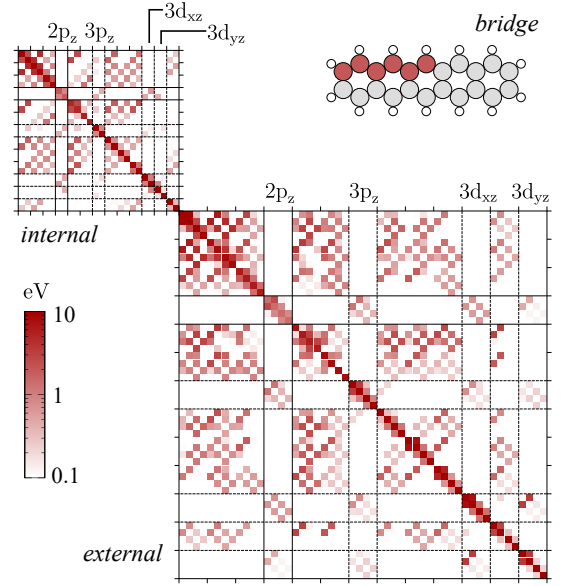}
    \end{center}
    \caption{Subdiagonalized blocks of $\Hm_S$ for representative internal and external carbon atoms 
    of the PA bridge (as indicated in the inset), sorted by energy. 
    The solid lines highlight the 4x4 and the 2x2 blocks of quasi-degenerate $\text{2p}_\text{z}$ LOs. 
    The color indicates the absolute value of the inter-orbital coupling 
    (or the LO eigenvalue, on the diagonal).}  
    \label{fig:PA_Hss}
\end{figure}

We subdiagonalize all carbon atoms and neglect the hydrogens. 
In the LCAO double-$\zeta$ basis we describe each carbon with $13$ basis functions. 
There are two different atomic sites with different energies: internal and external carbons, 
depending on whether they possess a C-H bond or only C-C bonds, respectively. 
It is convenient to analyze the Hamiltonian for these two kind of atoms separately, 
as shown in Fig.~\ref{fig:PA_Hss}. 
Next, we group together quasi-degenerate LOs~\footnote{The LOs are not exactly degenerate due to the different local chemical environments of the carbon atoms.} 
and then order the groups for increasing energy. 
Note that the LOs in the same group are centered at different atomic sites but possess the same local symmetry, 
e.g., the group closest to the Fermi level are $2\text{p}_\text{z}$ LOs. 
This can be verified by plotting the corresponding eigenvectors. 
By inspecting each matrix, we realize that there exist subsets of local orbitals that, 
to a first approximation, couple only within themselves  
(we verified that the inter-subset couplings are below $10^{-6}$~eV). 
We identify the one including the $2\text{p}_\text{z}$ as the subset of relevant LOs, 
which are able to accurately reproduce the transport properties of the junction 
around the Fermi level (see Sec.\ref{sec:PA}). 
A close inspection of the eigenvectors shows that the relevant LOs usually have 
$\text{2p}_\text{z}$, $\text{3p}_\text{z}$, $\text{3d}_\text{xz}$, and $\text{3d}_\text{yz}$ symmetry, 
as shown in Fig.~\ref{fig:PA_LOs}. 
Note that for internal and external carbon atoms, due to the different local chemical environment, 
these orbitals may have different energies and therefore occupy different positions in the ordered matrix. 

\begin{figure}[th!]
    \begin{center}
        \includegraphics[width=0.5\textwidth]{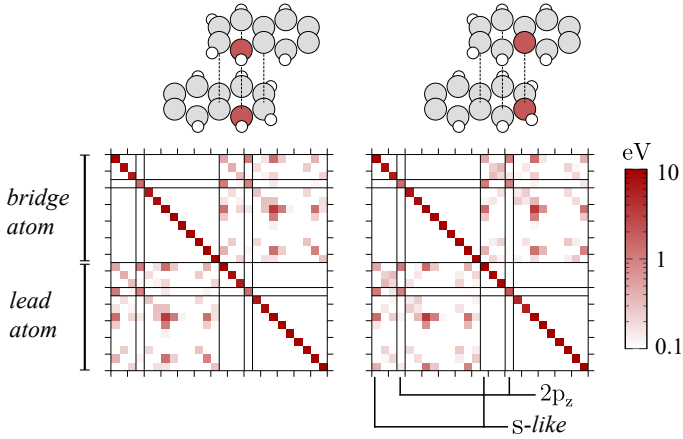}
    \end{center}
    \caption{Subdiagonalized blocks for representative pairs of carbon atoms of the $\pi$-stacked rings 
    of the PA junction (as indicated in the insets), sorted by energy for each atom. 
    The solid lines highlight the $\text{2p}_\text{z}$ LOs of each atom. 
    Due to the $\pi$-stacking, there is a sizable coupling between 
    $\text{2p}_\text{z}$ and $\text{s}$-like LOs, 
    which is negligible within the $\text{sp}^2$ planar structures. 
    The color indicates the absolute value of the inter-orbital coupling 
    (or the LO eigenvalue, on the diagonal).}  
    \label{fig:PA_Hss_stacked}
\end{figure}

For the sake of completeness, we also analyze the couplings between LOs 
for the $\pi$-stacked rings connecting the PA bridge to the rest of the junction. 
In Fig.~\ref{fig:PA_Hss_stacked} we show the subdiagonalized blocks of selected pairs of carbon atoms, 
with one atom belonging to the PA bridge, and the other to the PA lead 
(i.e., the non-periodic part of the lead, included in the scattering region). 
We identify a sizable coupling between $\text{2p}_\text{z}$ LOs, as expected, 
but also couplings between $\text{2p}_\text{z}$ and $\text{s}$-like LOs, 
which are negligible within $\text{sp}^2$ planar PA structures. 
One can expect similar couplings also between other pairs of carbon atoms 
e.g., atoms not stacked directly on top of each other along the $z$ direction. 
In the transmission function shown in Fig.~\ref{fig:trans_PA} 
we have neglected this limited number of couplings and retained only 
couplings between the relevant LOs, without reducing the quality of the results.

\subsection{BDA: Subdiagonalization with metal atomic contacts}
\begin{figure}[t!]
    \begin{center}
        \includegraphics[width=0.40\textwidth]{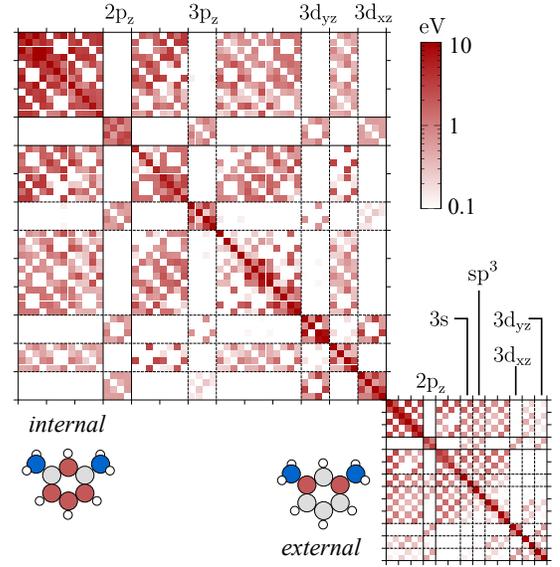}
    \end{center}
    \caption{ 
    Subdiagonalized blocks of $\Hm_S$ for the internal and external carbon atoms 
    of the 1,3-BDA molecular junction (as indicated in the insets), sorted by energy. 
    The solid lines highlight the 4x4 and 2x2 blocks of quasi-degenerate $\text{2p}_\text{z}$ LOs. 
    The color indicates the absolute value of the inter-orbital coupling 
    (or the LO eigenvalue, on the diagonal). } 
    \label{fig:meta_BDA_Hss}
\end{figure}

In this case, the Hamiltonian $\Hm_S$ includes only the six carbon atoms of benzene. 
It is convenient to look separately at the carbon atoms which directly bond to the NH$_2$ anchor groups 
and all others carbon atoms, which we label as external and internal, respectively 
(note the different notation with respect to the case of the PA). 
The reason is that due to the different local bond structure, the eigenvalues of the subdiagonalized 
Hamiltonian can change significantly, due to e.g., charge transfer between C and N. 
We consider explicitly that case of the 1,3-BDA, as shown in Fig.~\ref{fig:meta_BDA_Hss},  
but an analogous analysis can be done for the 1,4-BDA junction.

\begin{figure}[h!]
    \begin{center}
        \includegraphics[width=0.35\textwidth]{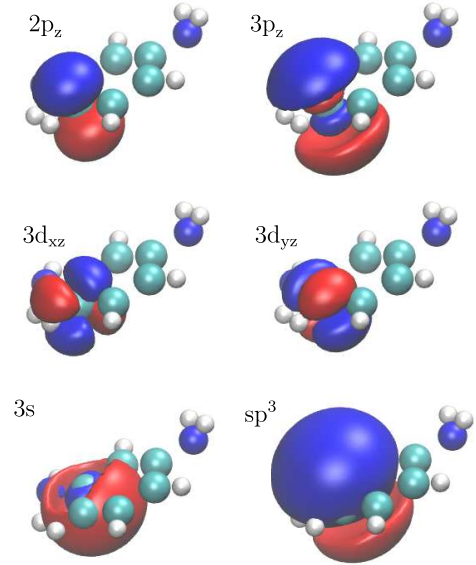}
    \end{center}
    \caption{Subset of relevant LOs for the carbon atoms of the 1,3-BDA molecule.  
    For the internal carbon atoms they have 
    $\text{3p}_\text{z}$, $\text{3d}_\text{xz}$, and $\text{3d}_\text{yz}$ character 
    while for external carbon atoms the $\text{3p}_\text{z}$ is replaced by 
    orbitals with $\text{3s}$ and $\text{sp}^3$-like symmetry instead. 
    Legend: C (cyan), H (white), N (blue).}
    \label{fig:meta_BDA_LOs}
\end{figure}

We group and order the LOs according to their energies as described for the case of the PA junction. 
We identify a set of relevant LOs which include the $\text{2p}_\text{z}$. 
For the internal carbon atoms the structure of the reordered $\Hm_S$ matrix 
is identical to that observed in the PA junction, whereas for external carbon atoms, there is coupling to an additional fourth LO. 
By plotting the corresponding eigenvector 
in Fig.~\ref{fig:meta_BDA_LOs} 
instead of the $\text{3p}_\text{z}$ LO, 
we identify two LOs with $\text{3s}$ and $\text{sp}^3$-like symmetry. 
This is likely to be a consequence of the slightly out-of-plane bonding 
with the amino group, and a change of hybridization character due to the C-N bond. 

At this point we note that, contrarily to the case of the PA junction where the matrix of the active space $\Hm_A$ is constructed from the set of relevant LOs only, here $\Hm_A$ 
also includes the LCAO orbitals of the N and Au atoms. 
The latter species were excluded from the subdiagonalization 
because it is beyond the scope of this article, 
but a similar analysis can be in principle performed for 
metallic surfaces as well.

\nocite{*}
\bibliographystyle{aip}
%

\end{document}